\documentclass{appolb}
\usepackage{amsmath,amssymb}
\usepackage{graphicx}

\DeclareTextFontCommand{\zapf}{\fontencoding{U}\fontfamily{pzd}\selectfont}

\def\ee{$e^+e^-$}               % e+ e- (annihilations...)

\def\pp{$pp$}

\def\nbar{\bar n}            % NB parameters: must be used in math mode
\def\Nbar{\bar N}            %   only, or it won't work!
\def\nc{{\bar n_c}}          % extra {} ensure subscriting it is ok.

\def\pt{p\kern -.2pt\lower 4pt\hbox{\tiny T}}    %works?
\def\mt{m\kern -.2pt\lower 4pt\hbox{\tiny T}}    %works?

\def\p0{P_0(\Delta y)}

\def\avg#1{\langle #1 \rangle}  % <n>:

\def\NF{\mathcal{N}_{\kern -1.9pt f}}
\def\NC{\mathcal{N}_{\kern -1.7pt c}}

\begin{document}
\title{Main results of a search on multiplicity distributions in 
pp collisions: is anybody afraid of a new class of hard events?%
\thanks{Presented at XXXIV INTERNATIONAL SYMPOSIUM ON MULTIPARTICLE DYNAMICS,
Sonoma County, California, USA, July 26-August 1, 2004}%
}
\author{\underline{R. Ugoccioni} and A. Giovannini
\address{Dipartimento di Fisica Teorica and INFN - Sezione di Torino\\
  Via P. Giuria 1, 10125 Torino, Italy}
}
\maketitle

\begin{abstract}
After an introduction on possible scenarios in the TeV energy region 
in \pp\
collisions, extrapolated from the knowledge of the GeV energy region,
attention is focused on the onset of a third class of events, harder than
the semi-hard and soft ones identified at SpS and Tevatron.
The expected features and signatures in multiplicity fluctuations,
forward-backward correlations and collision energy density
are discussed.
\end{abstract}

\newcommand{\st}[2]{{#1}_{\text{#2}}}
\newcommand{\stt}[3]{{#1}_{{#2},\text{#3}}}

\section{Introduction}

The weighted superposition mechanism (WSM) of two properly defined
classes of events explains several experimental facts
\cite{ARS:report}
(shoulder structure in $P_n$ vs $n$, quasi-oscillatory behaviour
of $H_q\equiv K_q/F_q$ vs $q$, energy dependence of forward-backward
multiplicity correlations)
which altogether characterise collective variables properties in high
energy \pp\ collisions and \ee\ annihilation.
In \pp\ collisions the two classes of events are the soft one (without
mini-jets) and the semi-hard one (with mini-jets); in \ee\
annihilation one
distinguishes between two-jet and three-jet samples of events.
%% Let us summarise the mentioned experimental facts \cite{ARS:report}:

%% \noindent 1) shoulder structure in the intermediate $n$-multiplicity
%% range of the $n$ charged particle multiplicity distribution (MD),
%% $P_n$, at top c.m.\ energies and in pseudo-rapidity intervals;

%% \noindent 2) quasi-oscillatory behaviour of the ratio of $n$-factorial
%% cumulants, $K_n$, to $n$-factorial moments, $F_n$, ($H_n = K_n/F_n$ in
%% the literature) after an initial sharp decrease towards a negative
%% minimum when plotted as a function of the order $n$ at different c.m.\
%% energies; 

%% \noindent 3) forward (F) --- backward (B) multiplicity correlation
%% strength, $\beta_{FB}$, energy dependence, with
%% \begin{equation}
%% 	\beta_{FB} = \frac{ \avg{(n_F -\nbar_F) (n_B - \nbar_B)} }{ 
%% 			         \left[\avg{(n_F -\nbar_F)^2}  \avg{(n_B - \nbar_B)^2}
%% 		           \right]^{1/2}} ,  \label{eq:FB.define}
%% \end{equation}
%% and $n_F$, $n_B$ the numbers of charged particles lying respectively 
%% in the forward and backward hemispheres, and $\nbar_F$ and $\nbar_B$
%% their corresponding average charged multiplicities.

It should be pointed out that the qualifying assumption of the WSM is
that the multiplicity distribution (MD) 
is described for each class of events in terms of the
Pascal, \ie, negative binomial (NB), MD; its characteristic parameters 
are $\nbar$, \ie, the average charged
particle multiplicity, and $k$ (linked to the variance $D^2
\equiv \avg{n^2} - \nbar^2$ by the relation $k=\nbar^2/(D^2-\nbar)$.)
This assumption leads to a sound
description of the experimental data.
The NB (Pascal) MD is well known in high energy physics 
and has been justified in the framework of QCD.

In extrapolating the WSM from the GeV to the TeV energy domain
\cite{combo:prd,combo:eta} for \pp\ collisions, 
it was found that, in the scenarios most
favoured by Tevatron results, the semi-hard component is characterised
by a decrease of the average number
of clans, $\st{\Nbar}{semi-hard}$, and by a corresponding increase of
the average number of particles per clan, $\stt{\nbar}{c}{semi-hard}$,
as the c.m.\ energy increases from 900 GeV to 14 TeV.
It seems that Van der Waals-like cohesive forces are at work among
clans. Somehow, clan aggregation is occurring and, accordingly, particle
population density per clan is expected to become larger as the c.m.\
energy increases.
Aggregation will of course be maximal when the average number of clans
reaches 1: under which conditions the
decrease of the average number of clans to one unit could be
extrapolated to 14 TeV?
furthermore, assuming that these conditions are verified at 14 TeV, are they
related to asymptotic properties the semi-hard component or are they the
benchmark of a new class of events, \ie, of a third component to
be added to the soft and semi-hard ones?
It should be pointed out that the onset of a third class of events in
terms of a second shoulder in $P_n$ vs $n$ is suggested in minimum
bias events in full phase-space also by Monte Carlo
calculations with Pythia version 6.210, with default parameters but using
a double-Gaussian matter distribution (so-called `model 4').

Coming to the first part of the above question, since one is forced to
exclude that a sudden decrease of the average number of clans to one
unit could be anticipated in the semi-hard component at 14 TeV c.m.\
energy (it would imply, quite unlikely, heavy discontinuities in
$\st{\nbar}{semi-hard}$ and $\st{k}{semi-hard}$ general behaviours),
it is proposed to consider
the following relation as the benchmark of a new class of events:
\begin{equation}
	\nbar = k\left(	  \e^{1/k} - 1 \right) ,
	  \label{eq:4prime}	
\end{equation}
which indeed implies $\Nbar=1$.

\section{Main properties of the new class of events}

We start by describing the general shape of the MD:
the KNO plot
reveals that events with 
multiplicities smaller than the average  are  numerous,
while events with multiplicities larger than the average 
are less probable although they extend to very large multiplicities.

Furthermore, the $k$ parameter, as well known, is related to the
integral of the two-particle correlation function. Thus one obtains
\begin{equation}
	\frac{ \st{\nbar}{III}^2}{ \st{k}{III}} = 
	   \int C_2^{\text{(III)}} (\eta_1,\eta_2) d \eta_1 d \eta_2  \gg 
		 \frac{ \st{\nbar}{semi-hard}^2}{ \st{k}{semi-hard}} ,
\end{equation}
because $\st{\nbar}{III} \gg \st{\nbar}{semi-hard}$ and
$\st{k}{III} \ll \st{k}{semi-hard}$;
this indicates that two-particle correlations in the new component 
are much larger than in the semi-hard component, a confirmation of
particle aggregation.
In addition, since $\st{k}{III}^{-1}$ controls $n$-factorial cumulant moments
behaviour at any order $n$,
$K_n^{\text{(III)}}$ is expected to be much larger than
$K_n^{\text{(semi-hard)}}$.

Finally, let us consider the strength of forward-backward
multiplicity correlations, 
$\stt{\beta}{FB}{III}$%, Eq.~(\ref{eq:FB.define})%
:
it was established in \cite{RU:FB} that
in the framework of clan structure analysis
\begin{equation}
   \stt{\beta}{FB}{III} \equiv \frac{ \avg{(n_F -\nbar_F) (n_B - \nbar_B)} }{ 
			         \left[\avg{(n_F -\nbar_F)^2}  \avg{(n_B - \nbar_B)^2}
		           \right]^{1/2}}
    =  \frac{ 2 \st{b}{III} \st{p}{III} 
		 ( 1- \st{p}{III}  ) }{ 1  - 2  \st{b}{III}  
		 \st{p}{III} (1-\st{p}{III}) },
\end{equation}
where $\st{b}{III} \equiv \st{\nbar}{III}/(\st{\nbar}{III} +
\st{k}{III})$, and $\st{p}{III}$ is the `leakage parameter', \ie, the
fraction of particles of one clan which do not leak to the hemisphere
opposite to where the clan was first emitted ($1/2 \leq \st{p}{III} \leq 1$).
Being for the third class $\st{\nbar}{III} \gg \st{k}{III}$, with
$\st{k}{III} \to 0$ one should have $\st{b}{III} \to 1$ and since
for $\st{\Nbar}{III} \to 1$ the corresponding leakage
controlling FB multiplicity correlations close to its maximum (leakage
parameter close to 1/2) one obtains that
$\stt{\beta}{FB}{III} \to 1$ and FB multiplicity
correlations in the third component are much stronger than in the
semi-hard class of events.
This appears to be an indication, 
in view of the extremely high virtuality and hardness
of these events, of a huge colour exchange process at parton level of
which strong FB multiplicity correlations are presumably the hadronic
signature. 

In conclusion, in this framework one should expect to see at 14 TeV in
\pp\ collisions three classes of events, each one described by a NB
(Pascal) MD or by one of its limiting values:

1. the class of soft events with $\st{k}{soft}$ constant as the
	c.m.\ energy increases;

2. the class of semi-hard events with $\st{k}{semi-hard}$ which
	decreases as the c.m.\ energy increases;

3. the class of hard events with $\st{\nbar}{III} \gg \st{k}{III}$
	and $0 \lesssim \st{k}{III} \ll 1$, \ie, $\st{\Nbar}{III} \simeq 1$

The total $n$ charged particle MD $P_n^{\text{total}}$ should
therefore be written as follows:
\begin{eqnarray}
	P_n^{(\text{total})} &=& \st{\alpha}{soft} 
	P_n^{\text{(NB Pascal)}} (\st{\nbar}{soft},\st{k}{soft})
	\nonumber\\ &+& 
	\st{\alpha}{semi-hard} 
	 P_n^{\text{(NB Pascal)}}(\st{\nbar}{semi-hard},\st{k}{semi-hard})
	 \nonumber\\ &+&
   \st{\alpha}{III} 
	 P_n^{\text{(NB Pascal)}}(\st{\nbar}{III},\st{k}{III}) ,
   \label{eq:6}
\end{eqnarray}
with $\st{\alpha}{soft} +\st{\alpha}{semi-hard} + \st{\alpha}{III}
=1$, where 
$\st{\alpha}{soft}$, $\st{\alpha}{semi-hard}$ and  $\st{\alpha}{III}$
are the weight factors of the three classes of events with respect to
the total sample of events.

%% \begin{figure}
%%   \begin{center}
%%   \includegraphics[width=0.45\textwidth]{plot_md3-combo.ps}
%% 	\hspace{0.05\textwidth}
%%   \includegraphics[width=0.45\textwidth]{plot_md_eta1.ps}
%%   \end{center}
%%   \caption[MD at 14 TeV with elbow]{$n$ charged particle
%%         multiplicity distribution $P_n$ expected at 14 TeV 
%% 				in full phase-space (left panel) and in $|\eta|<0.9$
%% 				(right panel) in presence
%%         of a third (maybe hard) component with
%%         $\Nbar_{\text{III}}=1$, showing one 
%%         shoulder structure and one `elbow' structure (solid line); the three 
%%         components are also shown: soft (dashed line), semi-hard
%%     (dash-dotted line) and the third (dotted band)}\label{fig:3}
%% \end{figure}

Assuming that at 14 TeV the third class of events is 2\% of the total
sample of events and that $\st{k}{III} \simeq 0.12$ and extrapolating 
$\st{\alpha}{soft}$ and $\st{\alpha}{semi-hard}$ from their behaviour
in the GeV energy range \cite{combo:prd} one gets in full
phase-space (FPS) the numbers in Table~\ref{tab:II}  (notice that
small variations of $\st{k}{III}$ below 0.12 in Equation
(\ref{eq:4prime}) give $\st{\nbar}{III} \gg 460$.)

In the pseudo-rapidity interval $|\eta| < 0.9$, assuming
that the clan is either (i)
spread out over all the phase-space or (ii) concentrated in $|\eta| <
0.9$,  the results in Table~\ref{tab:II} are obtained. 
%The superposition is shown in Figure~\ref{fig:3}.
For more details, see \cite{RU:NewPhysics,RU:NewPhysics:eta}.

\begin{table}
  \caption{Parameters of the three components at 14 TeV in full
  phase-space (FPS) and in a small rapidity interval (see text for  
  details.)}\label{tab:II}
  \vspace*{-0.2cm}
  \begin{center}
		\begin{tabular}[t]{|r|ccccc|}
			\hline
			{FPS}\vphantom{\LARGE H} &    \%   &
      $\nbar$   &   $k$ &  $\Nbar$ &  $\nc$\\
      \hline
		  soft       &    41  &   40  &  7     &  13.3 & 3.0\\
			semi-hard  &    57  &   87  &   3.7  &  11.8 & 7.4\\
			third      &    2   &  460  &  0.12  &  1    & 460\\
			\hline
			\hline
 			$ |\eta|<0.9$ &    \%   &
      $\nbar$   &   $k$ & $\Nbar$ &  $\nc$\\
			\hline
		  soft       &41  &    4.9  &     3.4   &   3.0   &   1.6\\
			semi-hard  & 57 &    14   &     2.0   &   4.2   &   3.4\\
			third (i)  & 2  &   40    &     0.06  &   0.37  &  110\\
			third (ii) & 2  &   460   &     0.12  &   1     &  460\\
			\hline
		\end{tabular}
	\end{center}
\end{table}

In addition, Bjorken formula \cite{Bj:energy} for the energy density,
\begin{equation}
        \varepsilon = \frac32 \frac{\avg{E_T}}{V} 
                \left.\frac{dn}{dy}\right\vert_{y=0},
\end{equation}
where $\avg{E_T}$ is the average transverse energy per particle, $V$
the collision volume and $dn/dy$ the particle density at mid-rapidity,
has been applied in order to compare its predictions on \pp\ collisions
with those on nucleus-nucleus collisions.
Parameters of the formula and results are shown in Table~\ref{tab:E}:
lacking general expectations for the
average transverse energy $\avg{E_T}$, we used for the soft component
the value measured at ISR and, in a conservative way, the value
measured by CDF for the other components (to be intended as a lower
bound, which leads to lower bounds for the energy density as well).
It should be noticed that
the energy density for the semi-hard component in our scenario at 
14~TeV is  of the same order of magnitude as that
found at AGS at 5.6~$A$GeV in O+Cu collisions;
the energy density for the third component in the spread-out
scenario is comparable with the value
recently measured at RHIC in Au+Au collisions; and
in the other extreme
scenario (high concentration) it is even larger, 
because of a large $dn/dy$, than the LHC
expectations for central Pb+Pb collisions ($\varepsilon\gtrsim 15$
GeV/fm${}^3$).
Of course our calculation of $\varepsilon$ is only indicative and
should be taken with caution:
although the use of Bjorken formula for \pp\ collisions as well as the
choice of parameters is rather doubtful, we consider our results quite
stimulating because they suggest the possibility that the same
characteristic behaviour of many observables seen at RHIC energies in
AA collisions could be reproduced at LHC in \pp\ collisions.

%% Bjorken formula for the energy density:\hspace*{1cm}
%% \begin{equation}
%% 	\varepsilon = \frac32 \frac{\avg{E_T}}{V} 
%% 	\left.\frac{dn}{dy}\right\vert_{y=0}
%% \end{equation}
%% where $\avg{E_T}$ is the average transverse energy per particle, $V$
%% the collision volume and $dn/dy$ the particle density at mid-rapidity.
%% Compare with
%% 	\begin{itemize}
%% 		\item AGS -- O + Cu -- $\sqrt{s_{NN}} = 5.6$ GeV -- 
%% 			$\varepsilon \approx 1.7$ GeV/fm${}^3$
%% 	\item RHIC (Phenix) -- Au + Au -- $\sqrt{s_{NN}} = 130$ GeV -- 
%% 		$\varepsilon \approx 4.6$ GeV/fm${}^3$
%% 	\end{itemize}
%% Results in Table~\ref{tab:E}.

\begin{table}
  \caption{Energy density and corresponding parameters for our scenarios and for
  Pythia Monte Carlo. The volume $V=\pi R^2\tau$ has been computed with
  proton radius $R\approx 1.1$~fm and formation time 
  $\tau\approx 1$~fm.}\label{tab:E}
  \vspace*{-0.3cm}
	\begin{center}
  \begin{tabular}[t]{|r|cccc|cc|}
		\hline
		  & soft & semi-hard & 
		   \multicolumn{2}{c}{(i) third (ii)}\vline &
		   \multicolumn{2}{c}{(i) total (ii)}\vline\\
    \hline
		$dn/dy$ &  2.5 & 7 & 20 & 230 & 10.8 & 19.2 \\
		$\avg{E_T}$ (MeV)&  350 & 500 & 500 & 500 & 500 & 500 \\
		$\varepsilon$ (GeV/fm${}^3$)& 0.4 & 1.6 & 4.7 & 54 & 2.5 & 4.5\\
		\hline
  \end{tabular}
	\end{center}
\end{table}

\section{Conclusions}

The reduction of $\st{\Nbar}{semi-hard}$ in \pp\ collisions with the
increase of the c.m.\ energy in the TeV energy region (second and
third scenarios in \cite{combo:prd}) led us to
postulate a third class of hard events to be added to the soft and
semi-hard ones, whose benchmark is $\st{\Nbar}{III} \simeq 1$, \ie,
$\st{\nbar}{III} \gg \st{k}{III}$ and
$\st{k}{III} \ll 1$ with $\st{k}{III} \simeq 0$.
It should be stessed that 
informations coming from deep inelastic scattering (DIS) do not help to
understand this behaviour, which is a peculiar property of
hadron-hadron scattering; indeed the clan size in DIS tends to have a
leptonic character (contrary to the average number of clans which
tends to be hadronic) \cite{AGLVH:4}.

The main properties of this new class of events were discussed and
predictions at LHC presented.
The extension of this search to nucleus-nucleus collisions is under
investigation.

%%%%%%%%%%%%%% journal name in italic, volume number in bold.
%%%%%%%%%%%%%% is the part (A,B...) part of the name? I say yes!

\end{document}